\documentclass[aps,float,prl,psfig]{revtex4}
\input epsf
\usepackage{graphicx}
\newcommand{\beq}{\begin{equation}}
\newcommand{\beqa}{\begin{eqnarray}}
\newcommand{\eeq}{\end{equation}}
\newcommand{\eeqa}{\end{eqnarray}}

\newcommand{\lsim}{\lesssim}

\newcommand{\vect}[1]{\mbox{\boldmath${#1}$}}
\newcommand{\lmk}{\left(}
\newcommand{\rmk}{\right)}
\newcommand{\lnk}{\left\{ }
\newcommand{\rnk}{\right\} }
\newcommand{\lkk}{\left[}
\newcommand{\rkk}{\right]}
\newcommand{\lla}{\left\langle}

\newcommand{\rra}{\right\rangle}

\newcommand{\vex}{{\vect x}}

\newcommand{\ven}{\vect n}

\newcommand{\ve}{{\vect e}}
\newcommand{\hf}{{\hat f}}

\begin{document}

\title{
Quest for
circular polarization of   gravitational wave background and orbits of
laser interferometers in space
} 
\author{Naoki Seto
}
\affiliation{Department of Physics and Astronomy, 4186 Frederick Reines
Hall, University of California, Irvine, CA 92697
}

\begin{abstract}
We show that  isotropic component of circular polarization of
stochastic  gravitational wave background can be explored by breaking
 two dimensional configuration of multiple laser interferometers for
 correlation 
 analysis. By appropriately selecting orbital parameters for the proposed BBO
 mission, the circular polarization degree $\Pi$ can be measured down to
 $\Pi\sim 0.08 (\Omega_{GW}/10^{-15})^{-1}(SNR/5)$ with slightly ($\sim 10\%$) sacrificing  the detection limit for  the total intensity $\Omega_{GW}$  compared to the standard plane symmetric configuration.   This might allow us to detect signature of parity violation in the very early universe. 
\end{abstract}
\maketitle

\section{introduction} 

Due to extreme penetrating power of gravitational waves, observation of
the waves may enable us to study the very early universe in a way that
cannot be achieved with other  methods \cite{Allen:1996vm}. Recently, follow-on
missions to the Laser Interferometer Space Antenna (LISA) \cite{lisa}
have been
actively 
discussed to directly detect stochastic gravitational wave background
from the early universe around $\sim 1$Hz. The primary target for the
proposed missions, such as, the Big Bang Observer (BBO) \cite{bbo} and
DECIGO
\cite{Seto:2001qf}, is the gravitational wave
background produced at inflation. With  various observational supports
for inflation, existence of the background is plausible, while it is
currently 
difficult to predict its amplitude.

Meanwhile, considering the facts that observation of gravitational waves
 will 
be a truly new frontier of  cosmology and our understanding of
physics is limited at very high energy scale, it would be quite
meaningful to prepare flexibly to various observational
possibilities. Actually,   detection of  something
unexpected with odd properties will be generally more exciting than
confirmation 
of something widely anticipated.  For this purpose it is preferable that we
can study gravitational wave background  beyond its simple spectral
information, and model independent approach  would be effective with
regard to 
the sources of the background.

One of  fundamental as well as interesting properties of the background is
its circular polarization. Circular polarization characterizes asymmetry
of  amplitudes of  right- and left-handed waves, and its generation
is expected to be closely related to parity violation (see {\it e.g.} \cite{Lue:1998mq}).
In a recent paper \cite{p1} it was shown that LISA can measure the dipole ($l=1$)
and octupole ($l=3$) anisotropic patterns of circular polarization of
the background in a relatively clean manner, but, at the same time, 
 LISA
cannot 
capture  its monopole ($l=0$) mode due to a symmetric reason.
Since our observed universe is highly homogeneous and isotropic at
cosmological scales, it would
be essential to have sensitivity to the monopole mode of 
circular polarization of cosmological background.  
 The  proposed missions like BBO or DECIGO are 
planed to use 
multiple sets of interferometers to perform correlation analysis to
measure the total intensity $\Omega_{GW}$ of cosmological  background by beating
out  detector noises with a long-term signal integration. The 
standard configuration of these missions is to put the multiple 
interferometers on a same plane.
This is advantageous to get a good sensitivity to the total intensity
$\Omega_{GW}$, as a larger spatial separation between interferometers
  results in reducing their
overlapped 
responses to gravitational waves \cite{Flanagan:1993ix}. However, with this plane symmetric
configuration, we are totally blind to the monopole mode of  circular
polarization, as in the case of LISA. This
means that even if the isotropic background is circularly
polarized by  100\%, we will not be able to discriminate its extreme
nature.  
In this paper we study how well we can detect the monopole mode of 
circular 
polarization by breaking
symmetry 
of the detector configuration.

\section{circular polarization}
In the transverse-traceless gauge, the metric perturbation due to gravitational waves is expressed by
superposition of plane waves as follows:
\beq
h_{ab}(t,\vex)=\sum_{P=+,\times} \int^{\infty}_{-\infty} df \int_{S^2} d\ven~
h_P(f,\ven) e^{2\pi i f (t-\ven \cdot \vex) } \ve^P_{ab}(\ven),\label{plane}
\eeq
where $S^2$ is the unit sphere for the angular integral, the unite vector
$\ven=(\sin\theta\cos\phi, \sin\theta\sin\phi,\cos\theta)$ is  
 propagation direction of the waves, and  $\ve^+_{ab}$ and $\ve^\times_{ab}$ are the bases for transverse-traceless tensors.  We fix them as
$\ve^+_{}={\hat \ve}_\theta \otimes {\hat \ve}_\theta- {\hat \ve}_\phi
\otimes  {\hat \ve}_\phi$ and $\ve^\times_{}={\hat \ve}_\theta \otimes
{\hat 
\ve}_\phi+{\hat  
\ve}_\phi \otimes {\hat \ve}_\theta$ where two unit vectors ${\hat
\ve}_\theta$ and ${\hat 
\ve}_\phi$ are defined in a fixed spherical coordinate
system, as usual.     The frequency dependence   is easily resolved by
Fourier transformation, and we omit its
explicit dependence for notational simplicity, unless we need to keep
it. 
We decompose the covariance matrix  
$\lla h_{P1}(\ven) h_{P2}^* (\ven') \rra$ $(P1,P2=+,\times)$
for two polarization modes as
\beq
\frac{\delta_{drc}(\ven-\ven')}{4\pi}\left( 
           \begin{array}{@{\,}cc@{\,}}
           I(\ven)+Q(\ven) & U(\ven)-iV(\ven)  \\
            U(\ven)+iV(\ven) & I(\ven)-Q(\ven)  \\ 
           \end{array} \right), \label{matrix}
\eeq
where $I,Q,U$ and $V$ are the Stokes parameters and are
real by definition. The 
parameters $Q$ and $U$ are related to linear polarization, and their
combinations $Q\pm iU$ have spin $\pm 4$ and  are expanded in terms of
the spin-weighted spherical 
harmonics ${}_{\pm4}Y_{lm}(\ven)$ defined for $l\ge4 $
\cite{p1}. Therefore, they do 
not 
have modes naturally corresponding to the monopole pattern.  The parameter
$I$ represents the total intensity, while the parameter $V$ shows
asymmetry of  amplitudes of  right- and left-handed waves. These
meanings become apparent with using the circular polarization bases 
$\ve_{R,L}=(\ve_+\pm i\ve_\times)/\sqrt{2}$ for which the expansion
coefficients become
$h_{R,L}=(h_+\mp ih_\times)/\sqrt{2}$. The parameters $I$
and $V$  can be expressed as
$
I(\ven)= \lla |h_R|^2+|h_L|^2\rra /2=\lla |h_+|^2+|h_\times|^2\rra /2 $ and
$V(\ven)= \lla |h_R|^2-|h_L|^2\rra /2=i \lla h_+ 
h_\times^*- h_+^* h_\times \rra /2
$.
They are spin 0 quantities and have monopole modes $I_{00}$ and $V_{00}$
for spherical harmonic expansions $I(\ven)=\sum_{lm}I_{lm}Y_{lm}(\ven)$
and 
$V(\ven)=\sum_{lm}V_{lm}Y_{lm}(\ven)$.  
In this paper we mainly discuss how to capture the monopole mode of circular polarization signal
$V_{00}$ with 
laser interferometers in space.

To begin with, we  summarize  basic aspects of LISA \cite{lisa}. LISA is formed by
three 
spacecrafts that nearly keep a regular triangle with its side length
$L=5\times 10^6$km. The geometric barycenter of the triangle moves on a
circular orbit around the Sun with radius $\sim1$AU. The detector plane made by the
triangle inclines to the 
orbital plane of the barycenter by $\sim 60^\circ$ with changing its
orientation. The triangle also rotates annually on the detector plane
(the so-called ``cartwheel motions''). From its six one-way relative frequency
fluctuations of the laser light, we can make three
Time-Delay-Interferometer (TDI) variables $\lnk A,E,T\rnk$ that cancel the
laser frequency noises  \cite{Prince:2002hp}. The detector noises
between these three 
variables are not correlated due to a symmetry of the data streams, and
two modes $\lnk A,E\rnk$ have  similar noise spectra  \cite{Prince:2002hp}.
At low frequency regime $\hf\equiv2\pi f/L\ll 1$, the responses of two
modes $\lnk A,E\rnk$ to gravitational waves can be  regarded as
those of two  simple L-shaped interferometers that measure differences of
two arm-lengths caused by passing gravitational waves, as shown in
figure 1. The 
$T$-mode is less sensitive to gravitational waves at $\hf\lsim 1$ and is
not 
important in this paper (see {\it e.g.} \cite{Seto:2005qy,Corbin:2005ny}). Since the monopole modes $I_{00}$ and $V_{00}$ are
our main concern and they are
invariant under rotation of a coordinate system, we only use the
coordinate  system
$XYZ$ fixed to the single system B1 as shown in figure 1. 
These basic aspects for LISA are essentially same for BBO
\cite{bbo}. But BBO is planed
to have a 
smaller arm-length $L=5\times 10^4$km ($\hf=1$ corresponding to 0.95Hz)
and use multiple systems 
(triangles) not only the first system B1 \cite{bbo} (see also \cite{Seto:2005qy,Corbin:2005ny}). The responses of $A$ and $E$
modes to gravitational waves  are written as
\beq
\lnk A,E\rnk =\int_{S^2} d\ven  \sum_{P=+,\times} h_P(\ven) F^p_{\lnk A
,E\rnk}(\ven,\hf), \label{res}
\eeq
and the pattern functions $F^P_A(\ven,f)$ for $A$ mode are expressed as
$F_A^+(\ven,f)=\frac12 (1+\cos^2\theta)\sin(2\phi)+O(\hf)$ and
$F_A^\times(\ven,f)=\cos\theta \cos(2\phi)+O(\hf)$ (see {\it e.g.}
\cite{Taruya:2006kq,Seto:2002uj}). The functions $F^P_E(\ven,f)$ for $E$
mode is given by replacing $\phi$ with $\phi+\pi/4$. Here we multiplied an
appropriate common factor proportional to some powers of $\hf$ so that the
pattern functions become the simple form at the low frequency limit
$\hf\to 1$ as presented 
above. This normalization is just for illustrative purpose and  not
essential for our study. Note that we have correspondences such as
$F_A^+\to F_A^+$ and $F_A^\times \to -F_A^\times $  at
order $O(\hf^0)$ for a
plane symmetric 
replacement 
$\theta \to \pi/2-\theta$ and $\phi\to \phi$. These are indeed  valid at any
order $O(\hf^n)$  (see {\it e.g.}  IV.C in \cite{Taruya:2006kq}), as we
can expect from simple geometric consideration.

With data streams $A$ and $E$ from  a single system B1 we can  make three meaningful combinations
$AA^*$, $EE^*$ and $AE^*$. 
The expectation values for a
combination $C$ by  the monopole modes $I_{00}$ and $V_{00}$ can be  written as 
\beq
\lla C(f)\rra=\lkk
\gamma_{I,C}(f)I_{00}(f)+\gamma_{V,C}(f)V_{00}(f)\rkk/5. 
\eeq 
 The overlap functions $\gamma_{\lnk I,V\rnk,C}$ show
the relative strength of inputs $I_{00}$ and $V_{00}$ to an observable
$\lla C\rra$ \cite{Flanagan:1993ix}. They are given by the pattern functions, such as,
$F_A^{+}(\ven,\hf)$ and $F_A^{\times}(\ven,\hf)$ . For example, the function
$\gamma_{V,AE'^*}$ for $C=AE^*$ is given by 
\beq
\gamma_{V,AE*}(f)=\frac{5}{4\pi}\int_{S^2} d\ven  \lkk i \lnk
F_A^+F_{E}^{\times*}-
F_A^\times F_{E}^{+*} \rnk  \rkk.
\eeq
Here we used the definition (\ref{matrix}) and equation (\ref{res}).
In a same manner the function $\gamma_{I,AA'^*}$ is given by replacing the above
parenthesis $[\cdots]$ with $\lkk F_A^+ F_{A}^{+*}+ F_A^\times
F_{A}^{\times*}\rkk$. The kernel $[\cdots]$ for $\gamma_{V,AE'*}$ at
order $O(\hf^0)$ is given by
\beq
-i\lmk\frac{1+\cos^2\theta}{2}   \rmk \cos\theta. \label{ae}
\eeq
This factor is decomposed only with dipole $(l=1)$ and octupole $(l=3)$
patterns 
\cite{p1}, and 
cannot probe the monopole $V_{00}$.  
This is because 
 responses of interferometers to incident
waves have an apparent symmetry   with respect to the detector plane,
and  this cancellation holds at any order 
$O(\hf^n)$ (see {\it e.g.}  IV.C in \cite{Taruya:2006kq}). 
From the same reason we cannot probe the mode $V_{00}$ with using self
correlation, such as $\lla AA^*\rra$ or $\lla EE^*\rra$.
We need independent data streams to capture the target $V_{00}$.  Note that the kernel for
$\gamma_{I,AE^*}$ becomes $8^{-1}(1-\cos^2\theta)^2\sin4\phi$ at
$O(\hf^0)$. It is 
written only with hexadecapole modes $(l=4)$  \cite{Taruya:2006kq,Seto:2004np}, and we have $\gamma_{I,AE^*}=0$ at
$O(\hf^0)$. As we see later, this is preferable to reduce the contamination
of $I_{00}$ to determine the target $V_{00}$ with using a combination that is a
refined version of $AE^*$.

Next we consider a second system B2 in addition to the first one B1
discussed so far. With the standard configuration of BBO, B2 is put at
position obtained by  rotating B1 around the $Z$-axis by $180^\circ$,
and arms of these two systems form a star-like shape on the $XY$-plane \cite{bbo,Seto:2005qy,Corbin:2005ny}.
But as discussed above, we can not capture $V_{00}$ with this
configuration due to the plane symmetry of interferometers.
Therefore, we study a simple case with breaking this symmetry. We
consider to put B2 (more precisely its barycenter) on a circular orbit
that has the same radius ($\sim1$AU) as B1, but its orbital plane is
inclined to that of B1 with an angle $\varepsilon=D/{\rm 1AU}\lsim 10^{-3}$ in
units of radian. Here the parameter $D$ is the maximum distance between
barycenters of 
B1 and B2, and its preferable scale is $\sim 10^5$km, namely the same
order as the arm-length $L$ of BBO, as we see later. The two orbits of the barycenters intersect twice per orbital
period $T_{orb} (\sim 1{\rm yr})$. In figure 2 their configurations are shown with viewing from their node. We neglect tiny misalignment of directions of two
detector 
planes of order $\varepsilon$, and only study effects caused
by their relative positions. By dealing with the rotation of detector
planes and the cartwheel motions mentioned earlier, we can follow the position of the B2's
barycenter on the moving $XYZ$-coordinate attached to B1. The
trajectory of B2's barycenter $(B_X,B_Y,B_Z)$ is given as 
\beqa
d_X&\equiv& B_X/L=\sqrt{3}d(\cos\omega \cos(\omega+\alpha))/2,\\
d_Y&\equiv& B_Y/L=\sqrt{3}d(\cos\omega \sin(\omega+\alpha))/2,\\
d_Z&\equiv& B_Z/L=d \cos\omega/2.
\eeqa
Here we have defined
$d\equiv {D}/L$, and the parameter $\omega=2\pi(t/T_{orb})$
 is the orbital phase of B1 around the Sun. In figure 1 we show the
trajectory of B2
for $(\alpha,d)=(0,0.24)$ as dotted curves.
The standard BBO configuration is recovered with putting $d=0$.
 The free parameter $\alpha$
determines the orientation the dotted curves around the $Z$-axis, and we
hereafter fix it at $\alpha=0$.

\begin{figure}
  \begin{center}
\epsfxsize=9.cm
\begin{minipage}{\epsfxsize} \epsffile{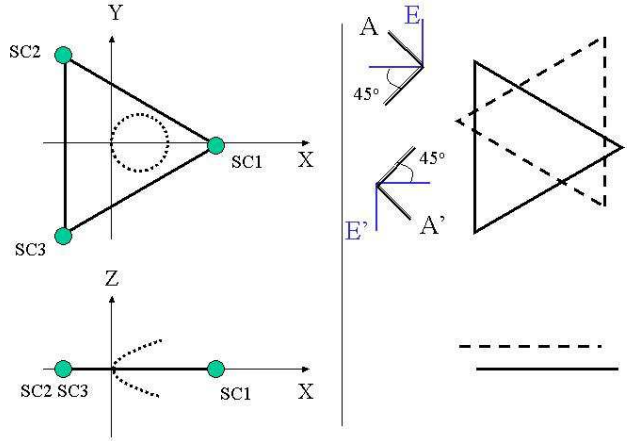} \end{minipage}
 \end{center}
  \caption{  Three spacecrafts (SCs) of a LISA-type system (B1) are shown with
 circles that are put on the $XY$-plane. The barycenter of three SCs is
 at the origin $(X,Y,Z)=(0,0,0)$.   The bottom figure shows the
 projection of the  structures to the $XZ$-plane. At low frequency limit the
 responses of two TDI  modes $A$ and $E$ can be effectively regarded as
 those of
 two L-shaped 
 interferometers with shown orientations. On the coordinate system $(X,Y,Z)$
 fixed to the first system B1,  the center of the second system B2 moves
 on the the dotted 
 curves for the specific parameter choice $(\alpha,d)=(0,0.24)$. A typical
 snapshot of B1 (solid line) and 
 B2 (dashed line) is shown
 on the right side.  The orientations of two effective L-shaped
 interferometers for $A'$ and $E'$ modes are also given.
 }
\end{figure}

\begin{figure}
  \begin{center}
\epsfxsize=7.cm
\begin{minipage}{\epsfxsize} \epsffile{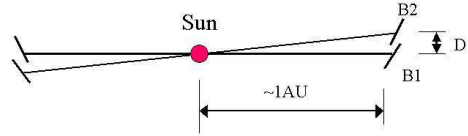} \end{minipage}
 \end{center}
  \caption{ Configuration of two orbital planes for B1 and B2
  seen from their node (orbital phase: $\omega=\pi/2$).  
 }
\end{figure}

We  define two TDI modes $A'$
and $E'$ made from B2 system in the same way as $A$ and $E$ from B1.
Now two modes $A'$ and $E'$  are not on the $XY$-plane (except for
$\omega= \pi/2,3\pi/2,\cdots$), and this introduces a phase shift $e^{-i\hf
d_Z n_Z}=1-i \hf d_Z n_Z+O(\hf^2)$ for their pattern functions
$F^P_{A',E'}$ (including information of position in the
$XYZ$-coordinate) from the previous ones 
$F^P_{A,E}$. When we take a  
combination $AE'^*$ (or almost equivalently $EA'^*$), this phase shift
generates a multiplier factor $\cos\theta(= n_Z)$ to 
 eq.(\ref{ae}) at order $O(\hf)$. Consequently, the combination
$AE'^*$ can  capture the monopole mode $V_{00}$ at $O(\hf)$, since we have
a  kernel proportional to $(1+\cos^2\theta)\cos^2\theta\ge 0$ for
circular polarization $V(\ven)$.
One the other hand, the combination $AA'^*$ (or equivalently $EE'^*$) can  be used to
detect the total intensity $I_{00}$  by the correlation technique, as for
the standard 
choice with $d=0$. But it is important to check how  the overlap function
$\gamma_{I,AA'^*}$  is reduced with taking finite distances $d\ne 0$.

We take a closer look at these aspects with  including all the higher order effects
$O(\hf^n)$.  Relevant overlap functions are numerically evaluated, and
some of the results are shown in figure 3.  
In this calculation we included not only the phase shift induced by the
relative  bulk
positions between B1 and B2, but also the effects by the finiteness of the
arm-length  $L$ \cite{Seto:2002uj}. 
As is well known for aligned interferometers, the function
$|\gamma_{I,AA'^*}|$ approaches 1 at the low
frequency limit. We have the following asymptotic profiles; 
$|\gamma_{I,AA'^*}|=1+O(\hf^2)$, $|\gamma_{V,AE'^*}|=2\hf |d_z| /3+O(\hf^3)$  and
$\gamma_{I,AE'^*} = O(d\hf^2)$.  The first nonvanishing term of  $\gamma_{I,AE'^*}$
 is determined only by  $d_X$ and $d_Y$. At $d=0$
corresponding to the simple traditional choice with putting B1 and B2 on
the $XY$-plane, we 
have identically $\gamma_{V,AE'^*}=0$ and the monopole $V_{00}$ cannot
be measured, as explained earlier. When we increase the separation $d$
(in figure 3: long-dashed curves $\to$ solid curves $\to$ short-dashed
curves), the combination $AA'^*$ loses sensitivity to the total intensity
$I_{00}$ from larger $\hf$. But the function $|\gamma_{V,AE'^*}|$ becomes
larger at small $\hf$ as shown by the asymptotic behavior, and the data
$\lla AE'^*\rra$ get better sensitivity there to the target $V_{00}$. In
figure 3 the 
results for 
$\gamma_{I,AE'^*}$ are presented as a reference to show potential
contamination of the total intensity $I_{00}$ for measuring the target
$V_{00}$ from  the data $\lla AE'^* \rra$. We do not go into this
effect. But, in many cases, it would be possible to estimate and
subtract this contamination relatively well, as the intensity $I_{00}$ is 
generally determined better than the target $V_{00}$ with using
observables such as $\lla AA'^*\rra$.  Note also that this
contamination might be somewhat reduced by adjusting 
free  parameters including  $\alpha$.

\begin{figure}
  \begin{center}
\epsfxsize=7.5cm
\begin{minipage}{\epsfxsize} \epsffile{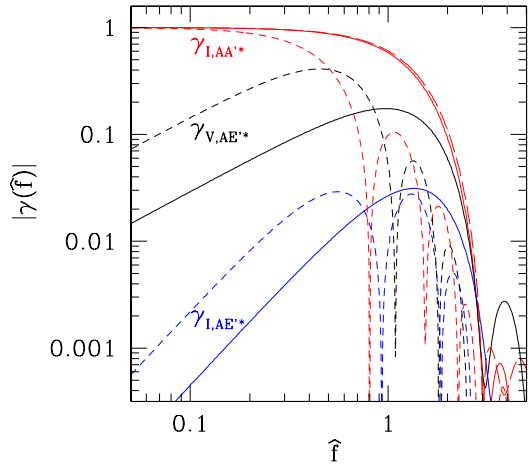} \end{minipage}
 \end{center}
  \caption{ Overlap reduction functions  for the unpolarized $I$ and
 circularly polarized $V$  
 modes for parameters $\omega=0.5$ and  $\alpha=0$. The long-dashed curve is
 result 
 with $d=0$ (traditional 
 choice) for which  we have $\gamma_{V,AE'*}=\gamma_{I,AE'*}=0$. The
 solid curves are for $d=1$, and the 
 short-dashed ones are for  $d=5$. The functions
 $|\gamma_{I,AA'*}|$  approach 1 at $\hf\to 0$, while the functions
 $\gamma_{V,AE'*}$ and   $\gamma_{I,AE'*}$ have asymptotic behaviors
 $O(\hf^1)$ and 
 $O(\hf^2)$ respectively. For   $\alpha=0$ we have
 $\gamma_{I,AE'*}=0$ (within numerical 
 errors) at $\omega=0$.
 }
\end{figure}

\section{correlation analysis}

We are now in a position to discuss how well we can estimate the
monopole $V_{00}$ of circular polarization of stochastic gravitational
wave background. We 
suppose that noises of relevant data streams $\lnk A,E,A',E'  \rnk$ are
not 
correlated and have identical spectrum $S(f)$ (as usually assumed for
BBO). Then the signal to noise ratio (SNR) for detecting $I_{00}$ with
using the combination $AA'^*$ is written as \cite{Flanagan:1993ix}
\beq
SNR_I^2=2 \lmk\frac{3H_0^2}{10\pi^2}  \rmk^2 \int_0^{T_{obs}} dt
\int_0^\infty df \frac{|\gamma_{I,AA'^*}(f,t)|^2\Omega_{GW}(f)^2}{f^6 S(f)^2}\label{snr},
\eeq
where we have used the familiar quantity
$
\Omega_{GW}(f)\equiv {4\pi^{3/2}
f^3 I_{00}(f)}/{3H_0^2} $ ($H_0$: Hubble parameter) and $T_{obs}$ is the
observational time. We also assumed that
the amplitude of the background is much weaker than the detector
noise. This corresponds to a situation when the correlation technique is
effective. The signal to noise ratio $SNR_V$ for the target $V_{00}$
with using the combination 
$AE'^*$ is given by replacing the simple amplitude $\Omega_{GW}(f)$  in eq.(\ref{snr}) with
the polarized one $\Pi \Omega_{GW}(f)$. Here the parameter
$\Pi\equiv V_{00}(f)/I_{00}(f)$ is the polarization degree and we
neglect its frequency  dependence in this paper.

\begin{figure}
  \begin{center}
\epsfxsize=7.5cm
\begin{minipage}{\epsfxsize} \epsffile{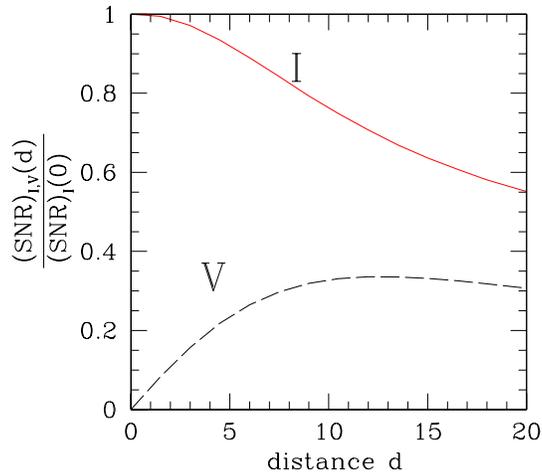} \end{minipage}
 \end{center}
  \caption{The signal to noise ratios for detecting $I_{00}$ and
 $V_{00}$ from combinations $\lla AA'^*\rra$ and $\lla AE'^*\rra$. The
 results are normalized with $SNR_I(0)$ for the simple orbital choice
 $d=0$. The shape of BBO noise curve is used and a flat spectrum
 $\Omega_{GW}=const$ is assumed. The polarization degree $\Pi$ is set at
 $\Pi=1$ for $SNR_V(d)$.  
 }
\end{figure}

The overlap functions $\gamma_{\lnk I,V \rnk,C}(f,t)$ depend on time $t$
through the change of the relative position between B1 and B2 as shown
in figure 1. With the designed noise spectrum $S(f)$ for BBO, we
numerically evaluate the integrals for $SNR_I$ and $SNR_V$ (with $\Pi=1$) as a function
of the maximum separation $d\equiv D/L$ between B1 and B2. We
take the observational time $T_{obs}$ as a natural number in units of
the orbital period $T_{orb}$, and assumed a flat
spectrum $\Omega_{GW}(f)=const$. In figure 4
ratios $SNR_I(d)/SNR_I(0)$ and $SNR_V(d)/SNR_I(0)$ are shown and these
are our central results. Since we have two relevant sets $\lnk AA'^*,
EE'^* \rnk $ and $\lnk AE'^*,
EA'^* \rnk $ for measuring $I_{00}$ and $V_{00}$ respectively, these
ratios 
can be effectively read as the results for the total network formed by B1 and
B2. When we increase the 
distance $d$, the sensitivity for the intensity $I_{00}$ decreases
monotonically due to 
 reduction of the overlap functions $\gamma_{I,AA'^*}$ as seen in
figure 3, but the sensitivity for  circular polarization increases for
 separation $d$ up to $\sim 12$. If we take $d=5$, the ratios are
$SNR_I(5)/SNR_I(0)=0.93$ and $SNR_V(5)/SNR_I(0)=0.24$.
This means that the detection limit for the intensity $\Omega_{GW}$  becomes
slightly ($\sim 
10\% $) worse compared with the simple conventional choice at 
$d=0$, but we can 
get  essentially new ``sensitivity'' to investigate circular
polarization of 
gravitational wave background.  For a  background with a flat spectrum
at
$\Omega_{GW}=10^{-15}$, BBO with $d=0$ has potential to detect it at
$SNR_I(0)=251$ by 10yr observation \cite{Seto:2005qy,note}. In other words its detection limit
is written as $\Omega_{GW,lim}=2\times 10^{-17}(SNR_I/5)(T_{obs}/10{\rm yr})^{-1/2}$. If we take
$d=5$ in stead of $d=0$, the limit becomes $\Omega_{GW,lim}=2.2\times 10^{-17}(SNR_I/5)(T_{obs}/10{\rm yr})^{-1/2}$
and we have the detection limit for circular polarization degree at $\Pi_{lim}=0.08 (\Omega_{GW}/10^{-15})^{-1}(SNR_V/5)(T_{obs}/10{\rm yr})^{-1/2}$.

The author would like to thank A. Taruya and T. Tanaka for comments, and
A. Cooray for various  supports.
This work was funded by McCue Fund at the Center for Cosmology,  
UC Irvine.




\begin{thebibliography}{DUM}
\bibitem{note}
This value is obtained with using parameters given in [3] for
 specification of BBO.
We should keep in mind that they do
not always represent its most up to date designed sensitivity.
\end{thebibliography}
\end{document}